\documentclass[12pt]{article}

\usepackage[pdftex]{graphicx}
\usepackage{cite}

\usepackage{multirow}%
\usepackage{amsmath,amssymb,amsfonts}%
\usepackage{amsthm}%
\usepackage{mathrsfs}%
\usepackage{xcolor}%
\usepackage{authblk}
\usepackage{hyperref}
\usepackage{microtype}

\usepackage{hyperref}
\usepackage{lineno}
\usepackage{slashed} 
\usepackage{SpinorNotation/Spinor} 
\usepackage{SpinorNotation/DiSpinor} 
\usepackage{SpinorNotation/ReOperators}
\usepackage{SpinorNotation/Clifford}
\usepackage{SpinorNotation/StandardModel}
 
\newcommand{\ds }{\displaystyle}

\begin{document}

\title{Polarized charged-current Drell-Yan process in {\tt ReneSANCe} generator}

\author[1]{S.\,Bondarenko}
\author[1,3,4]{Ya.\,Dydyshka}
\author[1,2]{L.\,Kalinovskaya}
\author[1,2]{A.\,Kampf} 
\author[1]{R.\,Sadykov}
\author[1,3,4]{V.\,Yermolchyk}

\affil[1]{\small Joint Institute for Nuclear Research, Dubna, 141980, Russia}
\affil[2]{\small Lomonosov Moscow State University, Moscow, 119991, Russia}
\affil[3]{\small INP, Belarusian State University, Minsk, 220006, Belarus}
\affil[4]{\small Dubna State University, Dubna, 141980, Russia}

\date{\today}

\maketitle

\abstract{
This work is devoted to validating the results obtained using the Monte Carlo generator {\tt ReneSANCe}. A comparison of differential cross sections, as well as single- and double-spin asymmetries, 
taking into account the polarization of initial states is presented.
}

\section*{Introduction}\label{sec:intro}

Complete one-loop electroweak radiative corrections to the charged-current Drell-Yan processes were presented for the case of longitudinal polarization of initial particles in the previous work \cite{Bondarenko:2024xzl}. This research contributes to a global next-to-leading order analysis of polarized parton distributions in proton-proton collisions at {\tt RHIC} \cite{RHICSPIN:2023zxx}.

Previously, a tuned comparison of the codes {\tt HORACE} \cite{CarloniCalame:2003ux,CarloniCalame:2006zq}, {\tt WGRAD2} \cite{Baur:1998kt} and { \tt SANC} \cite{Arbuzov:2005dd} for the unpolarized case was performed
and a good agreement was found
 (see section 4.4 of the proceedings \cite{TeV4LHC-Top:2007fwh}). 
Also, in paper \cite{Bondarenko:2024xzl} 
an agreement was also obtained 
at leading order with the works \cite{Zykunov:2001mn,Zykunov:2003gm} except in the polarized case.

In this brief note, we present the results of a comparison for polarized case  with those from works 
\cite{Gullenstern:1996pw} and with
\cite{Zykunov:2001mn,Zykunov:2003gm}. 

A comparison of the components of the  polarized differential cross section 
with those obtained using the {\tt SPHINX} program~\cite{Gullenstern:1996pw},
as well as of the single- and double-spin asymmetries in~\cite{Zykunov:2001mn,Zykunov:2003gm},
is presented.
It is important to note that the comparison was performed at the leading order and a good agreement  was found.

At the one-loop level, the 
works~\cite{Zykunov:2001mn,Zykunov:2003gm} do not include the subtraction of quark mass singularities.
Therefore, no comparison is possible in this case. Numerical results are obtained using the Monte-Carlo generator {\tt ReneSANCe} \cite{Bondarenko:2022mbi}.

The paper is organized as follows. In Section \ref{section1} the basic formulae for 
differential cross section and the polarized observables are given.
Section \ref{section3} is devoted to the presentation and discussion of the numerical results.
Our conclusions are drawn in Section \ref{Conclusion}.

\section{Spin-dependent observables}
\label{section1}

\label{section1}

Here, we present the basic formulae of calculating differential cross sections at one loop level.

\subsection{Hadronic level}

They take the following form at the hadronic level:
\begin{eqnarray*}
pp \to W^{+(-)} + X \to \ell^{+(-)}\nu_\ell(\bar{\nu}_\ell) + X,
\label{DYCC} 
\end{eqnarray*}
with $\ell= e,\mu$. 

The differential cross section is obtained by convolution with the corresponding parton distribution functions:
\begin{eqnarray*}
\label{sigpp}
d\sigma(\Lambda_1,\Lambda_2,s) =
\sum_{q_1 q_2}\sum_{\lambda_1\lambda_2}\int_{0}^{1}\int_{0}^{1} dx_1 dx_2
{f}_{q_1}^{\Lambda_1\lambda_1}(x_1) 
\times
{f}_{q_2}^{\Lambda_2\lambda_2}(x_2)\,
d\hat{\sigma}_{q_1 q_2}(\lambda_1,\lambda_2,\hat{s}), \nonumber
\end{eqnarray*}
where $\Lambda_i = \pm 1$ and $\lambda_i = \pm 1$ are the helicities of each proton and quark, respectively,
with $\hat{s} = x_1 x_2s$.

Parton distributions ${f}_{q_i}^{\Lambda_i\lambda_i}$ can be derived using unpolarized $f_{q_i}$ and longitudinally polarized $\Delta f_{q_i}$ PDFs:
${f}_{q_i}^{\Lambda_i\lambda_i} = \dfrac{1}{2}( f_{q_i} + \Lambda_i\lambda_i\Delta f_{q_i}).$

To derive longitudinally polarized cross sections, helicity amplitudes are calculated, and equation~(1.15) from~\cite{MoortgatPick:2005cw} is used.

\subsection{Partonic level}

At the partonic level, reactions take the following form:
\begin{gather*}
\begin{aligned}
\bar{d}(p_1,\lambda_1)  + 
  {u}(p_2,\lambda_2) \to 
l^+(p_3,\lambda_3)  + 
\nu_l(p_4,\lambda_4)~(+ \gamma(p_5,\lambda_5)), \\  
\bar{u}(p_1,\lambda_1)  + 
{d}(p_2,\lambda_2)\to  
{l}^-(p_3,\lambda_3) +  
\bar\nu_l(p_4,\lambda_4)~(+ \gamma(p_5,\lambda_5)),   
\end{aligned} 
\end{gather*}
where $p_i$ is momenta and $\lambda_i$ is helicity.  
As usual, the differential cross section is subdivided into five parts, as follows:
\begin{eqnarray*}
\hat{\sigma}^{\text{1-loop}}  =  \hat{\sigma}^{\mathrm{Born}} + \hat{\sigma}^{\mathrm{virt}}(\lambda) + \hat{\sigma}^{\mathrm{soft}}(\lambda, \omega)  
+ \hat{\sigma}^{\mathrm{hard}}(\omega) + \hat{\sigma}^{\mathrm{Subt}}, 
\label{loopxsec}
\end{eqnarray*}
where $\hat{\sigma}^{\mathrm{Born}}$ represents the contribution from the Born cross section, $\hat{\sigma}^{\mathrm{virt}}$ accounts for the virtual (loop) corrections, $\hat{\sigma}^{\mathrm{soft}}$ corresponds to the soft photon emission, and $\hat{\sigma}^{\mathrm{hard}}$ refers to the hard photon emission (with energy $E_{\gamma} > \omega$). Last two terms are defined using the soft-hard separator $\omega$, along with the auxiliary parameter $\lambda$ (a fictitious {\it photon mass} that regularizes infrared divergences). The special term $\hat{\sigma}^{\mathrm{Subt}}$ denotes the subtraction of collinear quark mass singularities. When all contributions to the cross section are summed, the result is free of infrared divergences. The partonic cross section is evaluated in the center-of-mass system of the initial quarks/antiquarks.
All contributions are obtained using the helicity amplitudes approach, with a sum over the helicities of all final state particles.

The differential cross sections, along with the single- and double-spin asymmetries, are compared to alternative codes. Here, some details of this comparison are discussed.  

\subsection{Polarized cross sections and asymmetries}

As mentioned in the Introduction, the comparison 
of the polarized differential cross section is carried out using the {\tt SPHINX} program. We compare the following components: $\sigma^{00}$, $\sigma^{++}$, $\sigma^{+-}$, 
$\sigma^{-+}$, $\sigma^{--}$, where $\sigma^{00}=\dfrac{1}{4}\left(\sigma^{++}+\sigma^{+-}+\sigma^{-+}+\sigma^{--}\right)$ is unpolarized cross section.

To define asymmetries, we introduce the following combinations of polarized components:
\begin{eqnarray*}
\Delta\sigma_{\mathrm {L}} &=&
\frac{1}{4}\left(\sigma^{++}+\sigma^{+-}-\sigma^{-+}-\sigma^{--}\right),\\   
\Delta\sigma_{\mathrm {LL}} &=&
\frac{1}{4}\left(\sigma^{++}-\sigma^{+-}-\sigma^{-+}+\sigma^{--}\right).
\end{eqnarray*}
Based on these combinations, the single-spin asymmetry ${A}_{\mathrm{L}}(\mathcal{O})$ and the double-spin  asymmetry ${A}_{\mathrm{LL}}(\mathcal{O})$ are defined as follows:
\begin{eqnarray*}\label{AL}\label{ALL}
{\text{A}}_{\mathrm{L}(\mathrm{LL})}(\mathcal{O})
 = \dfrac{{ d(\Delta\sigma_{\mathrm {L}(\mathrm{LL})})}/{\ds d \mathcal{O}}}
    {{\ds d\sigma}/{\ds d \mathcal{O}}},
\end{eqnarray*}
where $\mathcal{O}$ is some observable. The $z$ axis is directed along the momentum of the first proton. In the papers \cite{Zykunov:2001mn,Zykunov:2003gm}, the same conventions for beam polarization as in {\tt SPHINX} are used (relative to a $z^+$ axes direction).
 
\section{Numerical results}\label{sec:numres}
\label{section3}

\begin{figure}[!ht]
\begin{tabular}{cc}
    \hspace{-0.6cm}
    \includegraphics[width=0.5\textwidth]{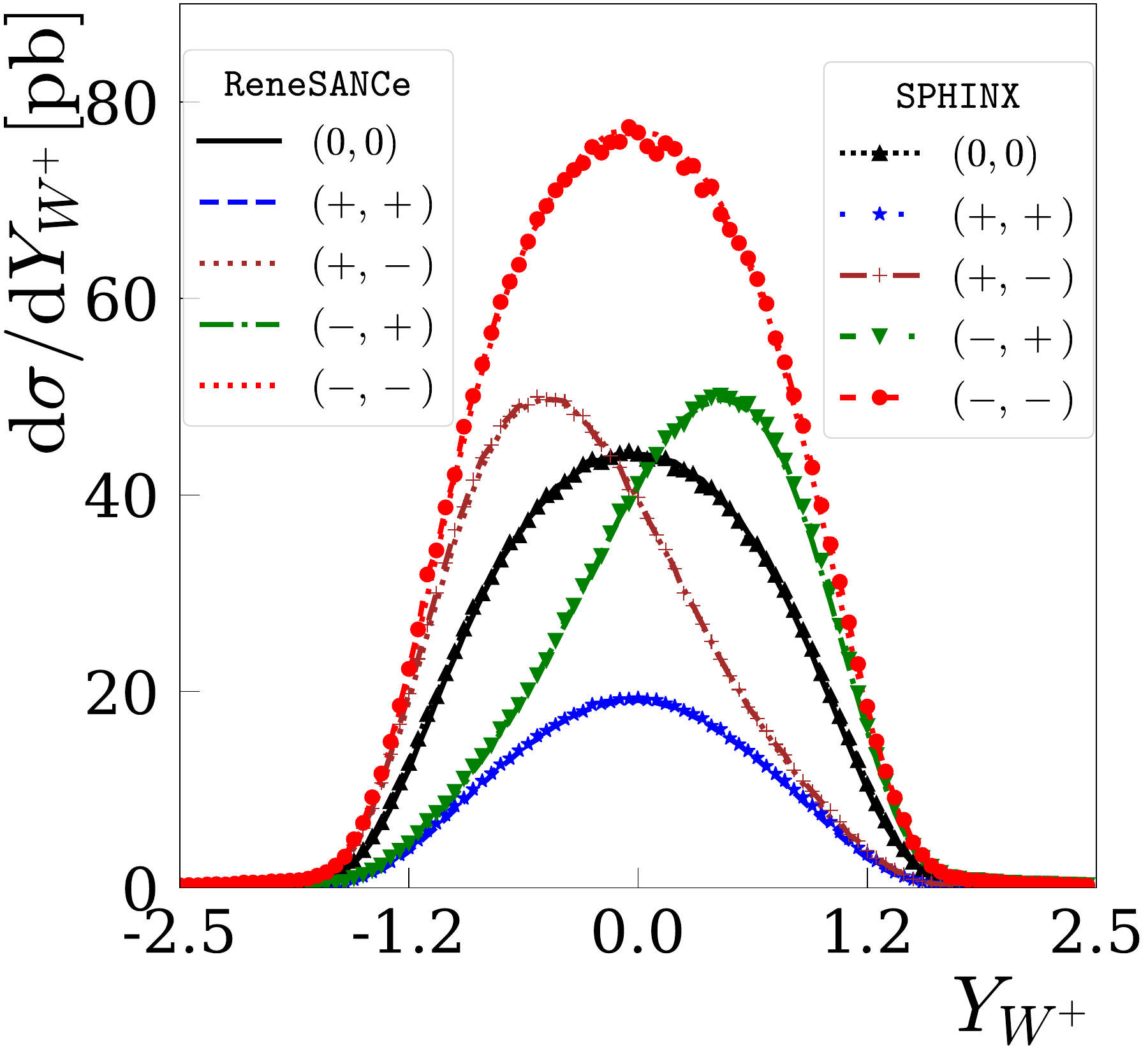} \hspace*{0mm}
    & \hspace*{0mm}\includegraphics[width=0.5\textwidth]{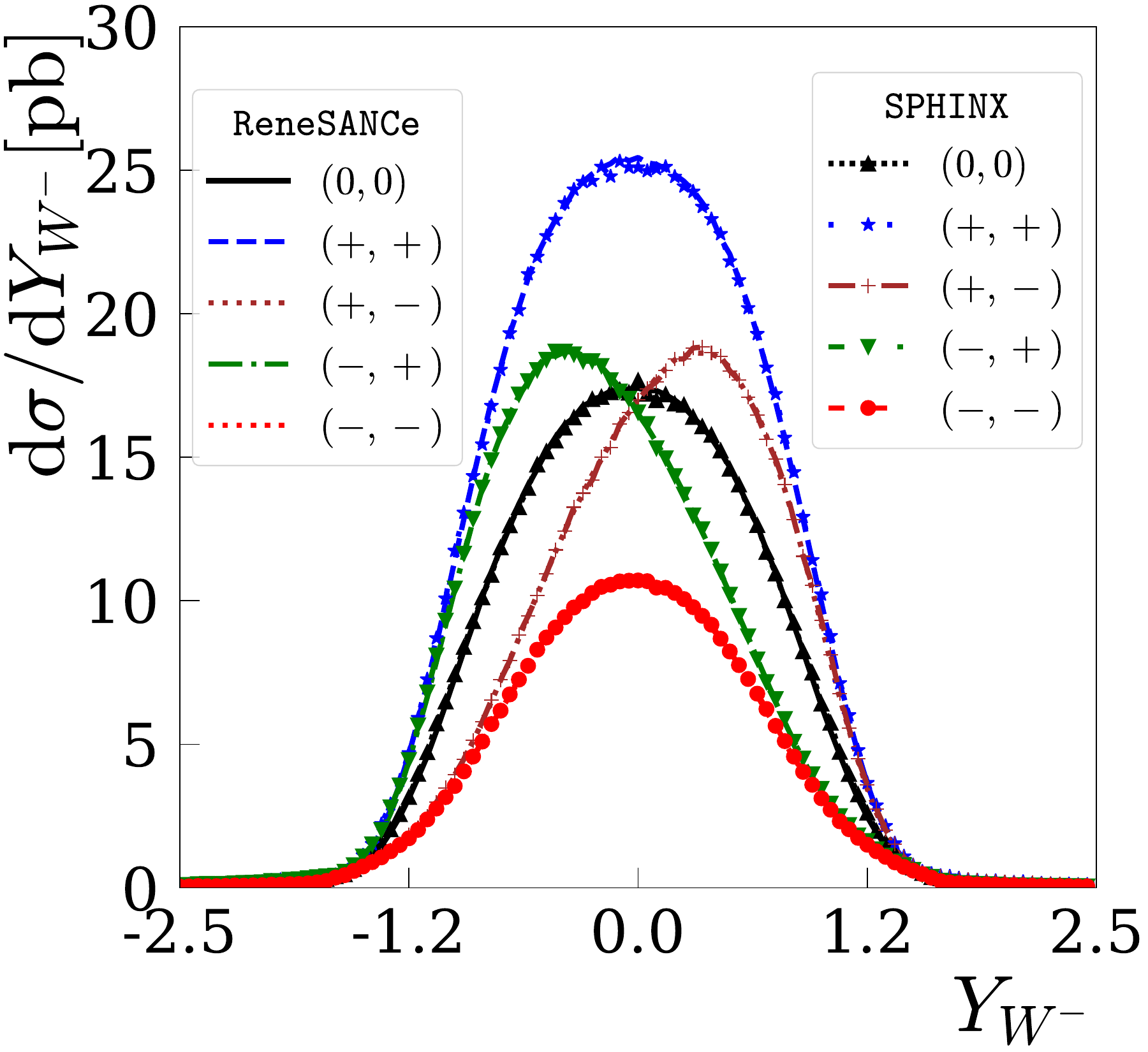}
\end{tabular}
    \caption{Results of the comparison differential cross sections as a function of rapidity for both W$^{+}$ and W$^{-}$ channels between {\tt ReneSANCe} and {\tt SPHINX}.\label{fig:CS}}
\end{figure}
\begin{figure}[!ht]
    \begin{tabular}{cc}
    \hspace{-0.6cm}
    \includegraphics[width=0.5\textwidth]{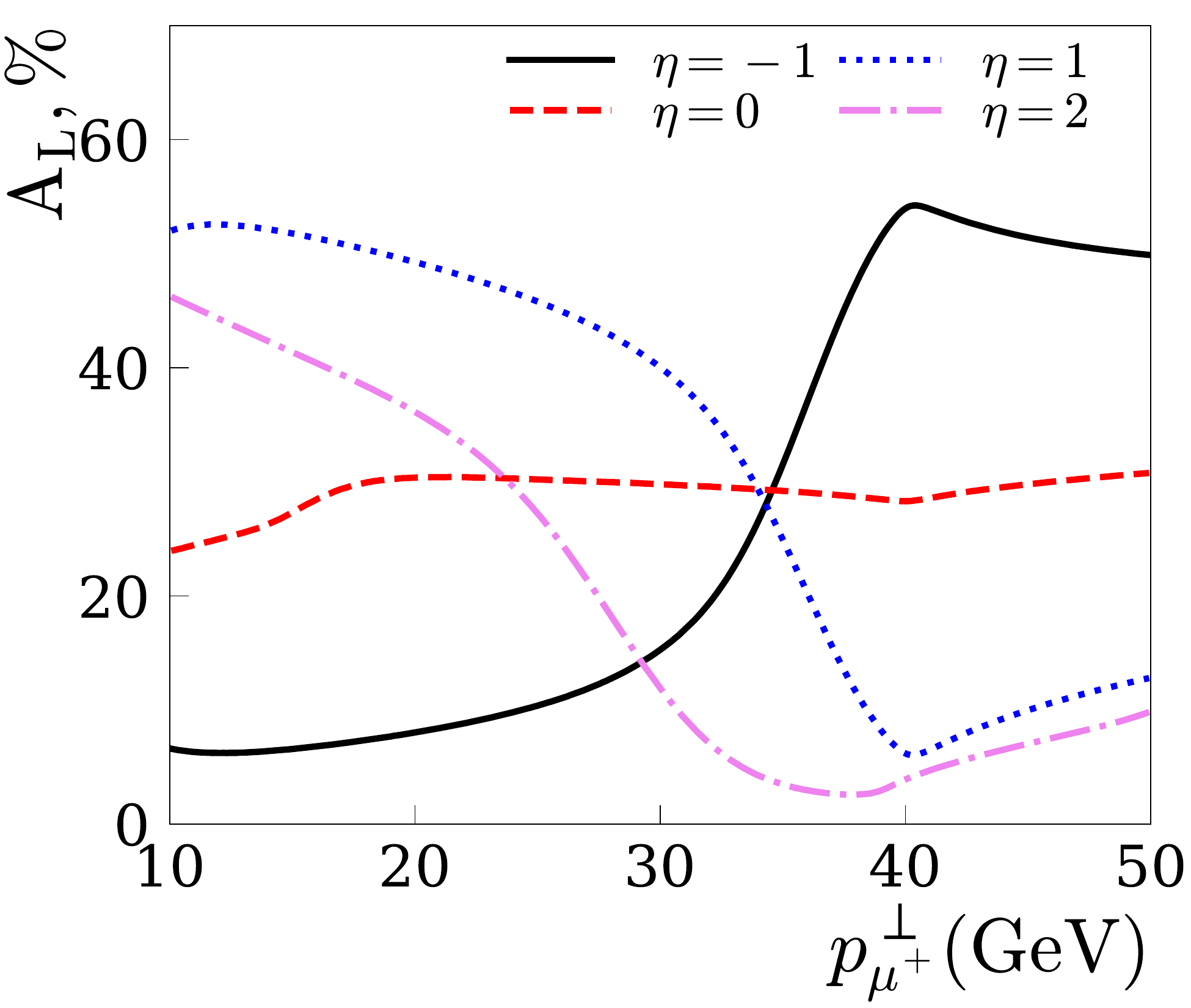} \hspace*{0mm}
    & \hspace*{0mm}\includegraphics[width=0.5\textwidth]{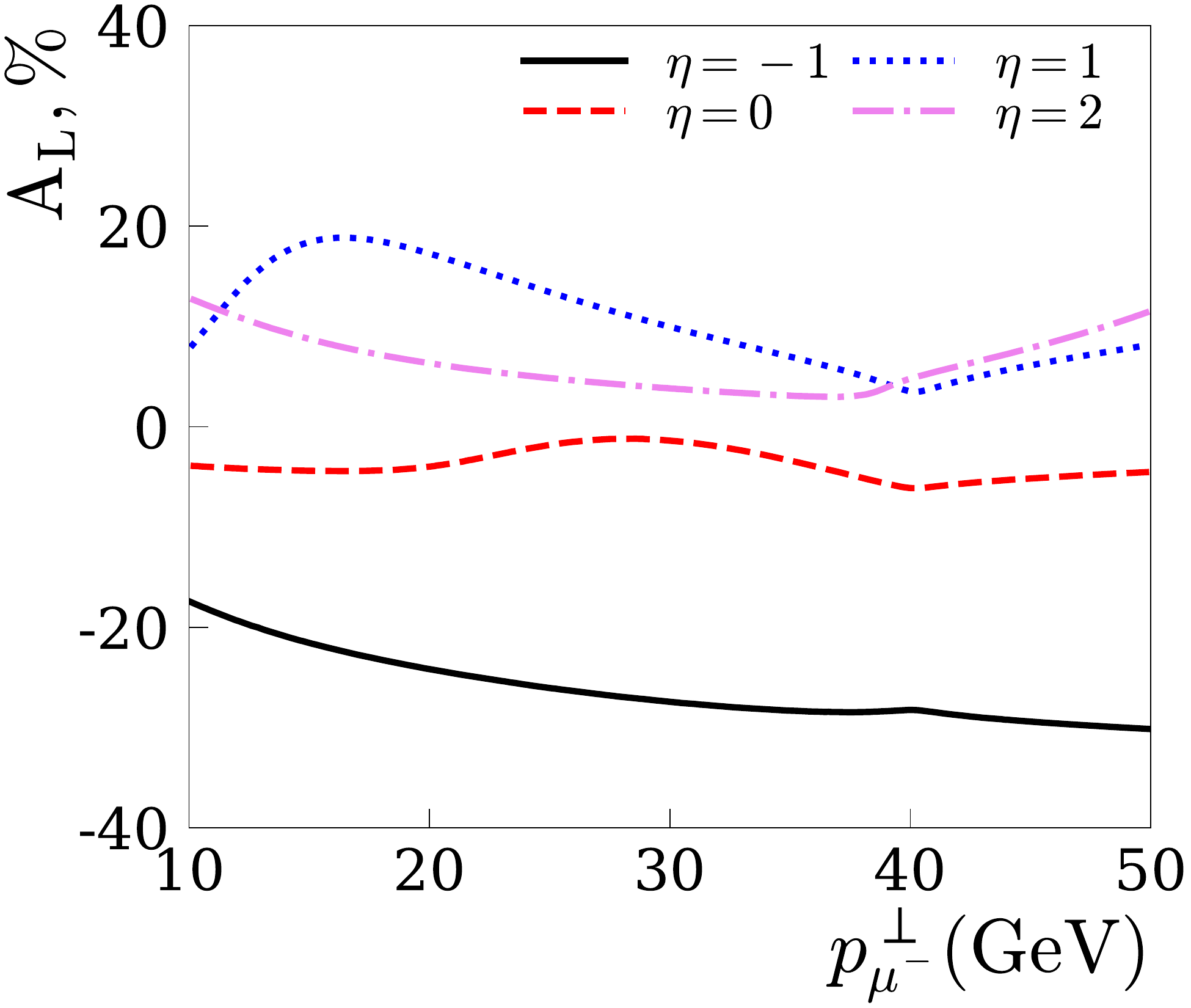}
    \end{tabular}
    \caption{Results for single-spin asymmetries as a function of transverse momentum for different values of pseudo-rapidity $\eta$.\label{fig:AL}}
\end{figure}  
\begin{figure}[!ht] 
    \begin{tabular}{cc}
    \hspace{-0.6cm}
    \includegraphics[width=0.5\textwidth]{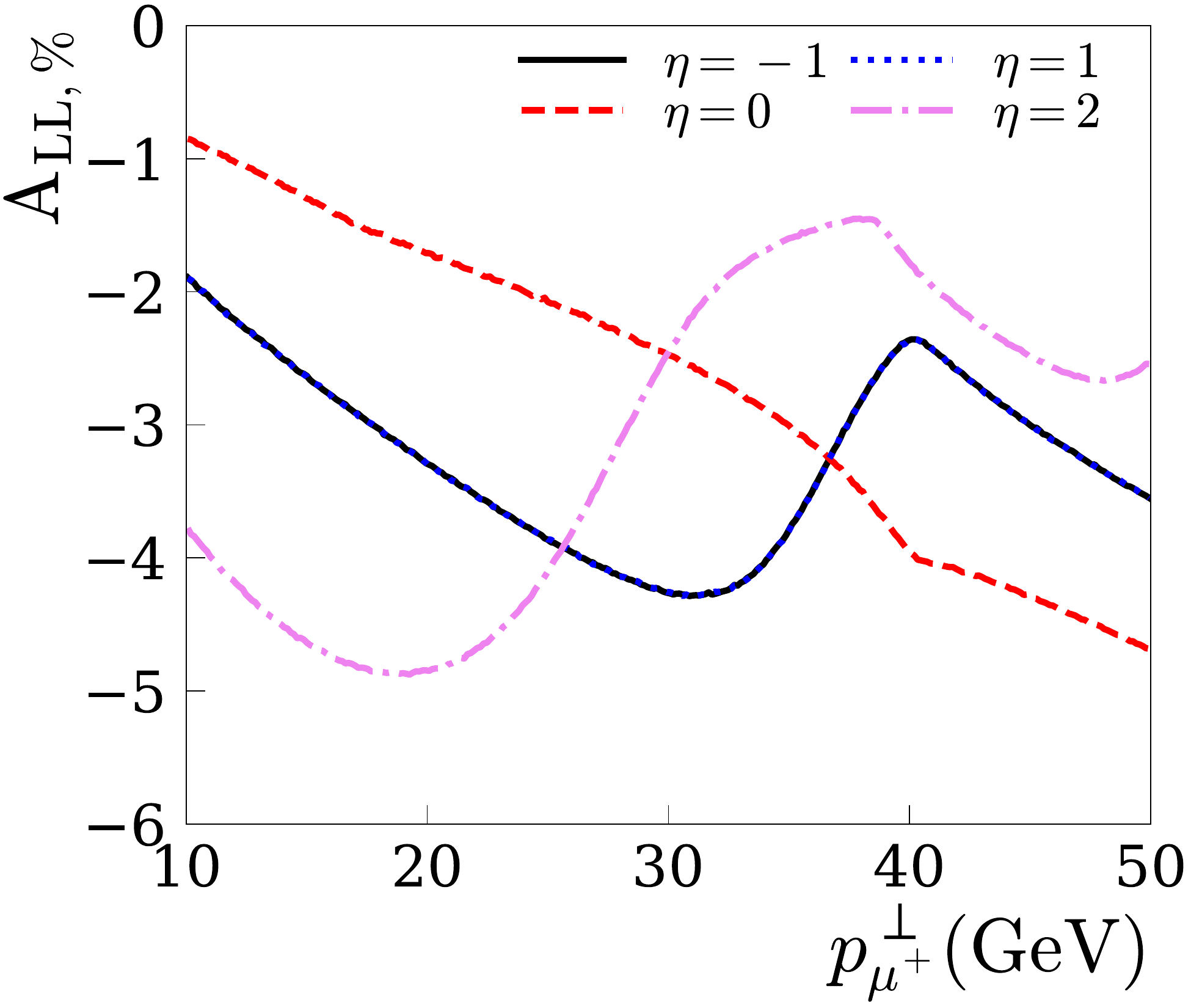} \hspace*{0mm}
    & \hspace*{0mm}\includegraphics[width=0.5\textwidth]{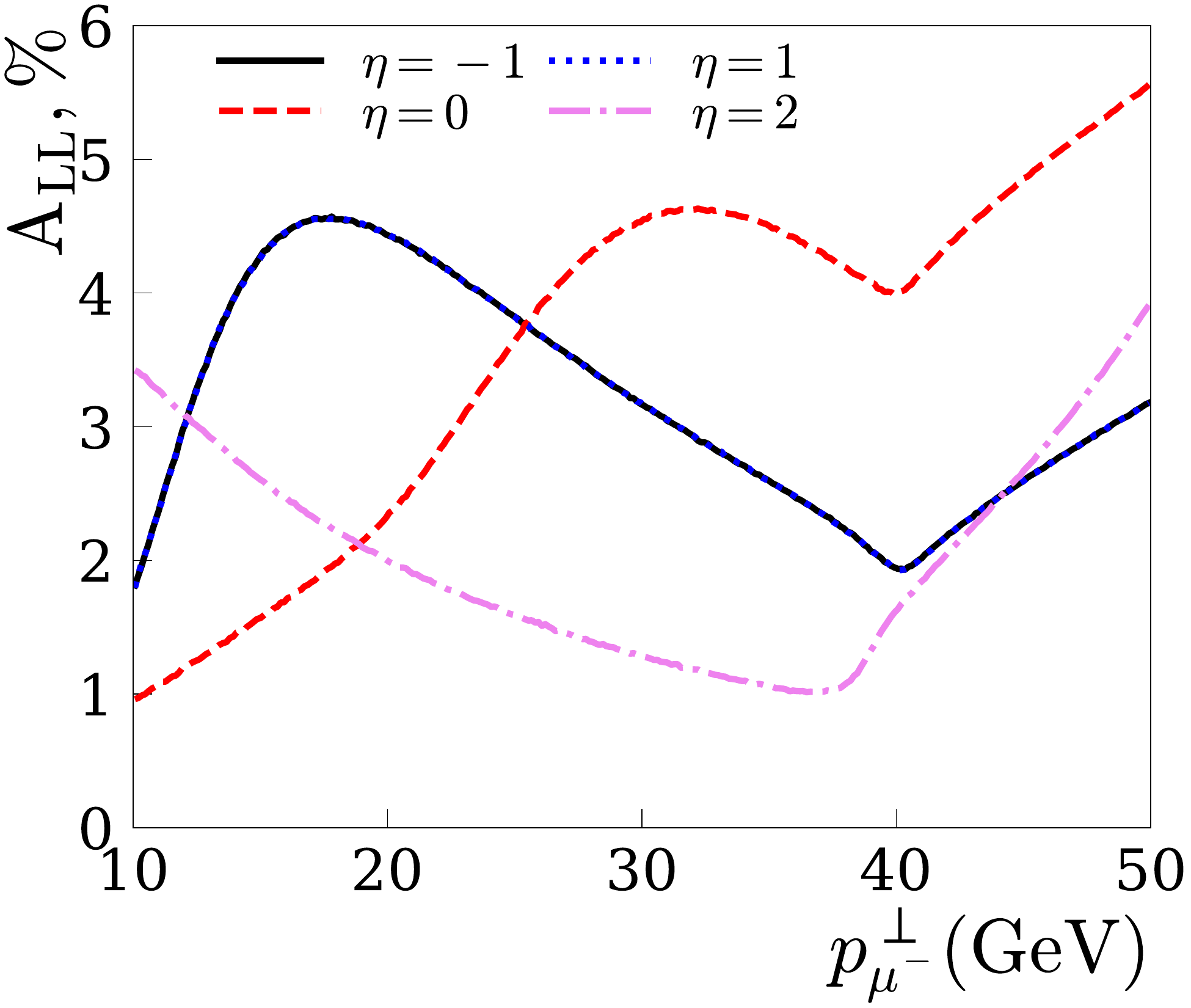}
    \end{tabular}
    \caption{Results for double-spin asymmetries as a function of transverse momentum for different values of pseudo-rapidity $\eta$.\label{fig:ALL}}
\end{figure}

In this section we present the results of our comparison of differential cross sections between {\tt ReneSANCe} and 
{\tt SPHINX} \cite{Gullenstern:1996pw}  
and for single- and double-asymmetries with works \cite{Zykunov:2001mn,Zykunov:2003gm}.

\subsection{Comparison of differential cross sections}

For a comparison, the set of input parameters is used as in \cite{Bondarenko:2024xzl}.
The cuts listed below are applied ($\ell = e, \; \mu$):
\begin{eqnarray*}
\label{Cuts}
&&W^{+}:\; p_\perp(\ell^+) > 2 \; \text{GeV}, \; p_\perp(\nu_\ell) > 2 \; \text{GeV},  M(\ell^+\nu_\ell) > 1 \; \text{GeV},  \\
&&W^{-}:\; p_\perp(\ell^-) > 2 \; \text{GeV}, \; p_\perp(\bar{\nu}_\ell) > 2 \; \text{GeV},  M(\ell^-\bar{\nu}_\ell) > 1 \; \text{GeV}.
\end{eqnarray*}

We use the PDF set {\tt NNPDF23\_nlo\_as\_0119} for unpolarized $f_{q_i}$ and the PDF set {\tt NNPDFpol11\_100} for longitudinally polarized $\Delta f_{q_i}$ parton distribution functions from the {\tt LHAPDF6} library, using a factorization scale of $\mu_F = M_{\ell\ell}$~\cite{ParticleDataGroup:2020ssz}.

Figure \ref{fig:CS} illustrates the differential cross sections as functions of rapidity $Y_{W^{\pm}}$. In this Figure, the results of {\tt ReneSANCe} are represented as curves without markers, whereas those of {\tt SPHINX} are shown with markers. As shown in Figure 1, the two codes show complete agreement across all combinations of initial state polarizations.

\subsection{Comparison of single- and double-asymmetries}

To compare asymmetries with those presented in works \cite{Zykunov:2001mn,Zykunov:2003gm}, we use the set of input parameters and kinematic cuts as described in these studies. However, some values of the input parameters used to obtain the results do not completely match those provided in the articles, as, for example, the width of $W$-boson and the sine of the Weinberg angle are not specified.

In Figures \ref{fig:AL} and \ref{fig:ALL}, the results for single- and double-spin asymmetries are presented. The $W^{+}$-channel is presented on the left, while the $W^{-}$-channel is shown on the right. These asymmetries are shown as function of transverse momentum for pseudo-rapidity values $-1$, 0, 1, and 2, which correspond to curves 1, 2, 3, and 5 from the \cite{Zykunov:2003gm}. To compare the results it is necessary to consider the dashed curves from the work \cite{Zykunov:2003gm}. A good agreement has been reached between {\tt ReneSANCe} and the mentioned above articles, despite the differences in some input parameters.
 
\section{Conclusions}
\label{Conclusion}
The polarized charged-current Drell-Yan process has been implemented in the Monte-Carlo generator {\tt ReneSANCe}. The distributions for single- and double-spin asymmetries were thoroughly compared with the results of the papers \cite{Zykunov:2001mn,Zykunov:2003gm} and the {\tt SPHINX} program and excellent agreement was obtained at the Born level. For the one-loop level, we obtained a result without double counting and independent of the quark masses. We have shown that the effect of radiative corrections on single- and double-spin asymmetries is negligible against the uncertainties of available polarized parton distribution functions.

\label{sec:funding}
\section*{Funding}
The research is supported by the Russian Science Foundation, project No. 22-12-00021.

\section*{Conflict of interest}
The authors of this work declare that they have no conflicts of interest.

\end{document}